**Authors**
Panayiotis Moutis [a], Spyros Skarvelis-Kazakos [b,1], Maria Brucoli [c]
**Affiliations**
[a] School of Electrical & Computer Engineering, NTUA, 9 Heroon Polytechniou, Zografou, 15780, Athens, Greece
[b] Faculty of Engineering and Science, University of Greenwich, Chatham Maritime, ME4 4TB, UK
[c] Arup, 13 Fitzroy Street, London, W1T 4BQ, United Kingdom

**Abstract**
Planned Communities (PCs) present a unique opportunity for deployment of intelligent control of demand-side distributed energy resources (DER) and storage, which may be organized in Microgrids (MGs). MGs require balancing for maintaining safe and resilient operation. This paper discusses the implications of using MG concepts for planning and control of energy systems within PCs. A novel tool is presented, based on decision trees (DT), with two potential applications: (i) planning of energy storage systems within such MGs and (ii) controlling energy resources for energy balancing within a PC MG. The energy storage planning and energy balancing methodology is validated through sensitivity case studies, demonstrating its effectiveness. A test implementation is presented, utilizing distributed controller hardware to execute the energy balancing algorithm in real-time.

**Keywords**: microgrids, decision trees, energy balancing, energy storage planning, intelligent control

## 1. Introduction

The microgrid (MG) paradigm is a Smart Grid application aims at facilitating more efficient management of loads and increased penetration of renewable generation in distribution networks [1]. Despite extensive trial applications, there has been no actual commercial deployment of the MG paradigm, but mostly retrofitting realizations [2-4], largely due to complex regulatory frameworks [4-6].

The planned community (PC) residential approach, such as "New Towns" in Hong Kong, has been gaining ground as a means of decongesting the overcrowded city centres. A PC constitutes a cell of energy consumption with various types of loads. Renewable Energy Sources (RES) or Distributed Energy Resources (DER), such as Distributed Generators (DG), Energy Storage and Controllable Loads, may be included in the PC energy portfolio. Management of all electrical resources and actors of a PC can be facilitated and optimized through approaches developed for MG [4]. The fact that the electrical network within the PC is in the ownership of the developer (i.e. private wires), makes the MG approach applicable at all levels of control and monitoring. Thus, PC MG present a unique opportunity to implement MG approaches from the network design phase, in contrast to the retrofitting applications deployed so far [6-9], which is the first contribution of this paper.

The PC may also be required to be connected to the medium voltage (or higher) and be served as an industrial customer or similar, due to its size. Hence, the PC MG can be defined as a prosumer – a possible participant in energy and/or ancillary services markets. A basic requirement in order to utilize the financial benefits of market participation is a concise daily and hourly operational energy balancing framework, such as the one developed in this work, which is the second and main contribution of this paper.

MGs can be operated either interconnected to the grid or as autonomous island systems [2,3]. Technical issues are associated with MG operation, in both modes [4]. These justify the need for advanced control methodologies, which take into account multiple variables, considerations and constraints. Two main control philosophies have been mainly applied to the MG paradigm: central and distributed. The advantages and disadvantages of each philosophy are outlined below [4]:

- The **Central** control strategy tends to be optimal due to its wider scope and is based on traditional power system management and control. However, communication infrastructure requirements are high, while loss of information may lead to loss of optimality or asset controllability. Advanced algorithms have been utilized, such as Heuristic Optimisation techniques (Genetic Algorithms and Ant Colony Optimization [4,41], Agent-based Potential Function method [42]) and Virtual Power Plant approaches [43].
- The **Distributed** control strategy is less demanding in infrastructure and caters for DER autonomy. However, it can be sub-optimal, complex DER communication and management protocols are required.

---

[1] Corresponding author. Tel.: 0044 1273 877352. E-mail: sskazakos@theiet.org. Present address: Department of Engineering and Design, University of Sussex, Brighton, BN1 9QT, UK

Intelligent agents have been mostly proposed for such control, comprising multiple distributed controllers, coordinating towards a common goal [4,10,44]; field trials of such approaches have proven their feasibility [11]. To a lesser degree techniques utilizing exclusively local information have also been discussed [45].

The density of stochastic residential loads, and possibly intermittent RES-based generation, introduces stochastic characteristics to the power balance in a PC MG [12-15]. These stochastic effects influence the economic scheduling of the reserves to be procured [16,17]. In addition, forecasting methods have accuracy limits, which may lead to deviations from day-ahead dispatching, introducing further uncertainty. Hence, the internal load and generation deviations from the expected values should be handled in an hour-ahead scheduling horizon (usual scheduling period in PSs [46]) based on very short-term forecasting.

The uncertainty of the loads and uncontrollable energy resources is a significant challenge for current energy balancing methods. Deviations, even from short-term forecasts, are always a possibility and it was shown in previous work by the authors that existing balancing methodologies such as intelligent control or centralized optimization methods cannot prevent that from happening [10].

This paper presents a novel intra-day internal energy balancing methodology for addressing both substantial load increase (or loss of local DER power) and substantial load reduction (or increase of local DER power) within a PC MG. The major novelty of the proposed methodology is that it can accommodate uncertainty both within the expected confidence intervals of the forecasts [12-15] as also beyond them [17], since it always comes up with a merit-order dispatch list that can be used as fall back in case of deviations from the near global optimum position. The cost of each dispatch as well as network technical constraints are taken into account. That way techniques suggesting increased reserves, which affect negatively PS economics [22-24], or curtailment of stochastic RES, which limits penetration of emission-free DG units [11], [24], [27] are avoided. A DT is used to extract each of two dispatches (one of increasing and one of decreasing total active power output) out of a large set of possible dispatches, analytically generated through Monte Carlo simulations. The Monte Carlo simulations generate the Learning Set (LS) of the DT according to the constraints, limits and required power reduction, while the DT extracts the dispatch that represents the most profitable (in terms of cost) solution for the PC MG owner/operator. Ensuring profitability of the corrective measures as described, avoids complex pricing methods [11] which introduce considerations for penalties in the day-ahead scheduling, as also the consideration for employing costly storage topologies [16],[28],[29].

The methodology was also developed as a tool for energy storage planning, since it can assess the installation characteristics of energy storage systems that most economically serve PC MG energy balancing deviations. This is the third major contribution of this paper.

The DT methodology is inherently centralized, which does not offer the benefits related to distributed approaches, as described above [4,10]. The final contribution of this paper is a proposed technique to decentralize the DT methodology, utilizing local PC MG controllers. Thus, the developed system incorporates the benefits of the DT methodology, but also takes advantage of the flexibility, extensibility and resilience of distributed control systems [4,10].

In Section 2 further details on the PC MG concept are offered along with some comments on the energy balancing problem of PC MG. The DT as a machine learning tool and the complete outline of the suggested methodology are analysed in Section 3. Simulated results are presented in Section 4, of a realistic PC MG case study assessing the proposed methodology, both as a storage planning investment appraisal tool and as an energy balancing control technique. A test implementation of the suggested methodology is presented in Section 5, while Section 6 concludes this work.

## 2. Planned Communities as microgrid entities – definition and incentives

In previous work, the authors have identified the main challenges for a PC installation to be voltage deviations beyond standardized limits, over-loading of electrical equipment and asymmetrical use of storage availability [18]. Hence, management of the available assets is required. However, the viability of investment in MG approaches is unclear; i.e. an assessment of the business case for MG applications to PC projects has to be performed.

RES DG units, especially wind, tidal and photovoltaic generation, are characterized by lower Operational Expenses (OpEx) compared to conventional plants and constantly decreasing Capital Expenses (CapEx). Hence, incorporating RES generation in the PC, even without considering any kind of incentive, is by itself profitable for the end customers' energy consumption. On the other hand adding batteries to any installation, despite being a step towards improved resilience, can be particularly expensive, due to their relatively high CapEx and short lifecycle [19]. In the UK energy market, Short-Term Operating Reserve (STOR) can be aggregated from multiple actors [20]. Utilising a battery storage system in a PC as a STOR contributor will greatly reduce its actual cost per kWh, while linking it with RES will offer additional value, due to mitigation of RES intermittency. It is assumed that the PC MG operator would have an obligation to follow a specific hourly schedule, due to its participation in energy or ancillary services markets, or because it may be closely controlled by the local system operator. Thus, any deviation from the day-ahead schedule will have to be dealt with internally or through the imbalances market [21]. Covering deviations internally may require committing the available storage reserve capacity; hence if STOR is called upon the PC MG, there will be – at least – a missed opportunity cost. Hence, the size of the storage system needs to be determined during system planning, as it affects PC MG economics.

The uncertainty of RES and demand has been proven in the literature [12-15]. According to [12] the averaged root mean square (RMS) error over a year for various day-ahead forecasting methods spans between 2.9-3.7%. Nevertheless, the point error during peak times can be considerably higher (±15%). In this paper, steep load changes have been modelled according to [13]. Wind power shows the highest error rates regarding its day-ahead forecast, at a normalized mean absolute error over a yearly period for the day-ahead forecast ranging between 2.4-9% [14]. The point errors can be as high as ±15% and under extreme cases as high as ±40%. Cloud shading is responsible for considerable deviations from the day-ahead forecasts of photovoltaics. The monthly average root mean square error of the day-ahead forecast for a photovoltaic plant is around 3.5%, however the point errors can deviate down to -35% and sometimes -50%. These high error rates occur because of partial clouding during the peak hours of photovoltaic power (around noon) [15]. <u>Hence, forecast deviation scenarios can be expected, with ±15% deviation of the load, deficit of photovoltaic generation down to -35% and a range of ±15% of wind power excursions</u>. Power system operators and current research suggest that the deviations should be dealt with through increased power reserves [22-24], curtailment of excess RES generation [25], [26], load shedding when DG is lost [11], [24], [27], active power conditioners [28] and storage devices [29]. The above measures affect power system economics, while RES curtailment discourages further penetration of sustainable generation.

Assuming that the RMS error of the forecasts follows a normal distribution, none of the above events should fall in the $\mu \pm 2\sigma$ confidence interval, but should only occur outside that interval. Hence, on a yearly basis, no more than 402 hourly deviations of each event (~4.6% of the year) can be considered. The events are assumed independent and their occurrence following a uniform distribution, i.e. there should be a total of 700 different instances of all possible combinations of the above events (load, PV and wind event each combined with none, one or both of the others). Some of these deviations cancel each other, thus require no measure to be taken. The PC MG will also ignore and not cover for deviations that are smaller than ±3.3% [12] from its day ahead schedule and only respond to deficits and excesses of power more than that. The ±3.3% was chosen as the smallest range of expected load forecast RMS error, since the system is considered primarily as a consumer.

## 3. Decision Tree (DT) methodology

Given that deviations cannot be known in advance, the PC MG operator/owner needs to have an economically near global optimum set of dispatch actions prepared for either power deficit or power excess. However, an analytical solution to energy balancing may not be possible, due to the random nature of both the loads and the RES-based generators installed in the PC MG. This is a probabilistic constrained non-linear optimization problem. The stochastic nature of certain variables of the dispatches (i.e. RES-based DG units of the PC MG) implies that a specific optimal dispatch may be impossible to realize, due to available power being below the confidence intervals initially assumed. In addition, adding up all balancing actions over a time period can define the average usability of the storage capacity installed, thus informing system planning. In this paper, a data-mining technique is proposed; DT has been considered to be the most suitable tool for the dispatch problem, as in [24, 30], as well as for energy storage capacity planning. DTs, unlike other machine learning approaches, can be used as an "off-the-shelf" data mining tool with only two weaknesses: (a) inability to extract linear relations among attributes, (b) poor response to conditions of the problem not present in the LS (low predictive power). The way that the proposed technique is developed and used makes both weaknesses irrelevant [24,30].

### 3.1. Decision Trees

A DT is a tree structure which extracts rules from a LS of pre-classified data [31]. In power system studies, DTs often process data of binary sorting, i.e. True-False, Safe-Unsafe, etc. Each internal node splits the available subset in two parts (children nodes) on a single attribute. If the subset of a child node is pure enough with respect to one of the classes, it is declared terminal, otherwise it is further split. Purity is assumed to be the percentage of "True" elements in the subset. Conventionally, the left child complies with the split criterion of the parent node, e.g. in the DT of Fig. 1, node 2 includes the subset of the LS for which A1≥8.2 containing no elements classified as "True".

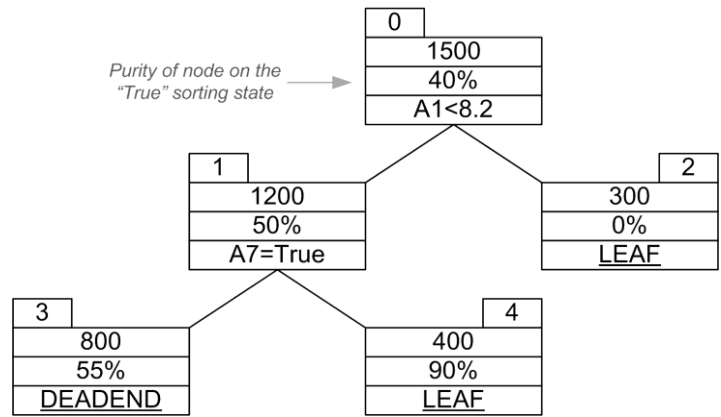

Fig.1. Example of a univariate, binary DT.

A terminal node can be either a leaf (acceptable purity and not split any further) or a dead-end (not acceptable purity and not split any further). For each leaf, the path leading to the root can be written in the form of if-then-else statements, which can be used as rules; e.g. the rules given from the DT of Fig.1 are:

*Rule (i): if* (A1≥8.2) *then* FALSE and *Rule (ii): if* (A1<8.2) *and* (A7=FALSE) *then* TRUE

The following characteristics of DT are important:
(a) The stop criteria define if the rules of the DT can be "generalized" or if the DT has been over-fitted to the LS. The split selection methodology defines whether "enough" rules were produced by the DT [32]. The stop criteria can either be a maximum number of iterations or an assessment of whether the information gain of splitting further any of the available nodes of the DT is sufficient. The second approach is considered suitable for the stochastic problem at hand [47], thus the $X^2$ criterion is used [37].
(b) The DT may not perform satisfactorily on unseen cases (generalization ability) [33].
(c) When, for a single LS, there exist multiple DTs describing the knowledge problem, then the DT with the highest generalization ability should be selected [33]. This may also apply to the case when several similar LSs can be used to build a DT for the same knowledge problem.

### *3.2. DT-aided energy balancing methodology for a planned community microgrid*

The energy balancing methodology will have to prepare schedules covering both for an excess and a loss of power. These two scheduling problems have been previously addressed separately in [34] and [24]. In this paper, the techniques are extended and combined to a single control strategy and adjusted to the specific parameters of a PC MG. Every hour *t* of the day, two sets of schedules for the hour *t+1* will be provided: one for increasing the PC MG total active power (supposing a loss of generation or excessive load have occurred) and a second for reducing it (supposing the DG has been unexpectedly increased or the load was less than the forecasted). The sets of schedules will be the rules extracted by DTs. The framework for the generation of the required LSs will be outlined.

#### *3.2.1. Power Deficit/Excess and Islanding*

Based on the day-ahead hourly scheduling of the PC MG, the maximum power deficit considered is either the loss of the largest active power injection of one (or more) DG unit(s) connected to a single bus of the PC MG or the largest possible positive load deviation according to the framework discussed in Section 2. Likewise, the maximum active power excess to be covered is either the largest possible negative load deviation or the largest possible positive DG deviation simultaneously occurring for all of the DG units of the same type (due to their proximity to the PC MG) on the basis of the analysis in Section 2. For the PC MG, it is crucial to balance generation and loads when it is forced into islanding from the interconnected power system. Depending on whether the PC MG is primarily a consumer or a producer of active power, part of its loads or part of its generation has to be curtailed, respectively, after its disconnection from the power system, to maintain a seamless operation.

#### *3.2.2. LS Evaluation of the DT for the planned community microgrid Energy Balancing Provision Schedules*

The cost of the balancing dispatch for each incident is taken as shown in (1), which is actually the cost function of PC MG; i.e. (1) calculates the revenue from selling power to the loads of the PC MG and any excess to the PS, minus the cost for buying energy from the grid, the operation cost of the units and/or the cost of curtailing loads according to demand response contracts signed with the PC MG loads.

$$\text{Dispatch Profit} = -\rho_{E,t+1}E_{t+1} + \rho_{R,t+1}R_{t+1} + \rho_{L,t+1}Load_{t+1} - \sum_{i \in S_{DG}}\left[C_{DG,i,t+1}\left(P_{DG,i,t+1} + R_{DG,i,t+1}\right)I_{i,t+1} + SC_{DG,i,t+1}J_{i,t+1}\right] - \sum_{i \in S_{int}}C_{int,i,t+1}R_{int,i,t+1} - \sum_{i \in S_{str}}C_{str,i,t+1}P_{str,i,t+1} \quad (1)$$

where $t+1$ denotes the hour-ahead scheduling horizon, $\rho_E$, $\rho_R$ and $\rho_L$ are the prices of energy, spinning reserve and retail energy, respectively, $E$ is the energy sold/bought by the PC MG to the market, $R$ is the sum of reserves of the DGs of the PC MG, $Load$ is the load served within the PC MG, $C$ represents the cost function of each DG, interruptible load (subscript *int*) or storage device (subscript *str*, both charging/bying and discharging/selling accordingly), $SC$ is the start-up or shut-down (accordingly) cost of each DG, $I$ and $J$ are binary variables denoting the operation and the start-up of each DG, respectively. The profit is calculated as the sum of the revenues of energy ($E_{t+1}<0$) and reserves sold to the market by the PC MG and the energy supplied to the load within the PC MG, minus the sum of the operating costs of the DG units, of the Demand Side Management (DSM) and of any start-up costs incurred during the hour. The technical constraints considered are the operational limits of the units, voltage and line loading limits, while the level of availability of the stochastic sources is based on the forecasting tools discussed earlier [12-15]. All other characteristics of the actors within the PC MG (battery storage, CHP, etc) affecting their operational cost are modelled according to the literature given earlier [19,38,42].

The LS evaluation is subject to voltage and loading constraints, as given in [35], [36]. From this point on, the most profitable schedules are favoured, i.e. after using (1) to calculate the balancing cost of every dispatch in the LS, certain values are selected separating the top-X% of the costs, e.g. top-30%, top-20%, top-10%, etc.; let each of these values be called **cost threshold**. There are two reasons why various cost thresholds (instead of just one, the most profitable) are examined. Firstly, for just one threshold, the method might yield a dispatch so specific that could suffer from reliability issues and possible loss of RES availability upon realization. Secondly, certain dispatches of high profit might account for technically non-feasible solutions due to technical constraints [35], [36]. Hence, the various balancing dispatches in the LS are evaluated as acceptable/true if both their cost is higher (more profitable) than the cost threshold and no technical constraints are violated.

The generation of the LSs of both DTs (one for each balancing requirement) is carried out through a Monte Carlo simulation around both the pre-disturbance state of the PC MG and the dispatch of the balancing among the DGs of the PC MG. In detail:
- The load of all buses between the load of hour $t$ and its forecast for hour $t+1$ is randomized.
- The assumed lost active power injection, plus its un-committed reserve, are randomly dispatched among the DG units of the PC MG, taking into account technical minima and RES availability accordingly.

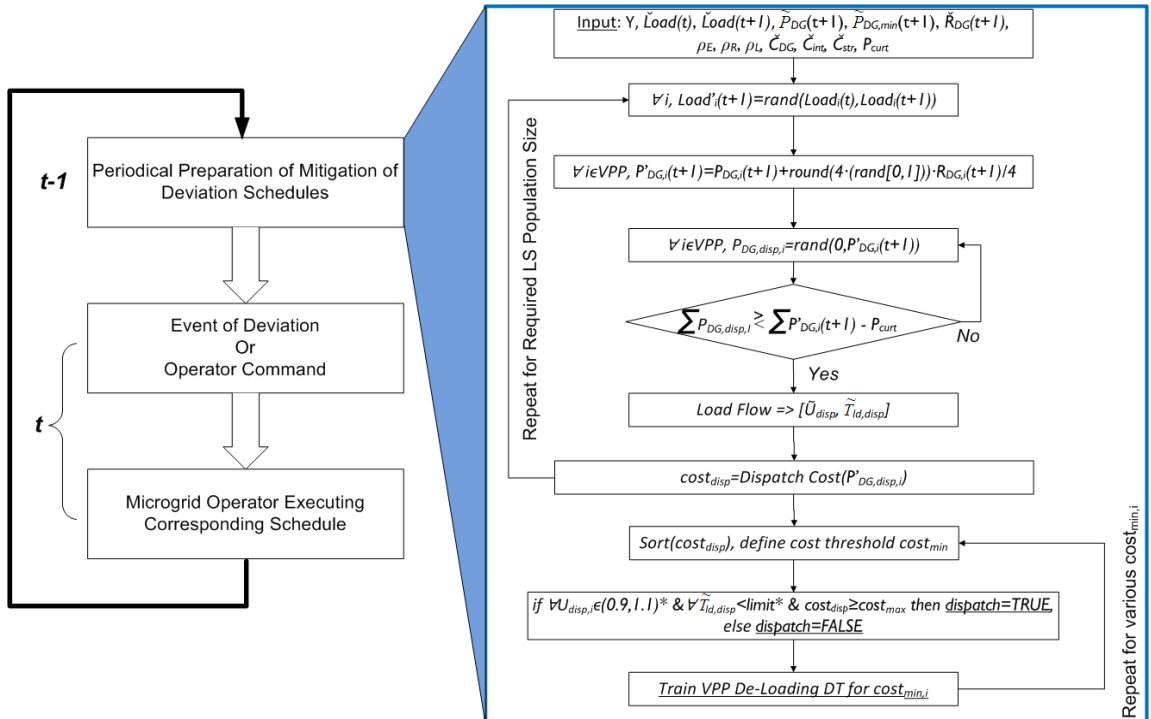

Fig. 2. Flowchart of the suggested DT-based energy balancing and storage planning methodology, *Y*: admittance matrix of the PS at hand, *Load(t)*: vector of the current loading of buses, *Load(t+1)*: vector of the forecasted loading of buses for hour $t+1$, *Pcurt*: required reduction/increase of active power of PC MG, rest as of *(1)*

### 3.2.3. Training of the DTs

The LS at the root and its subsets in the subsequent terminal nodes of the DT are split based on Shannon's entropy. More specifically, the normalized average mutual information gain C is calculated for every possible split in all terminal nodes of every step of the DT training. The split that yields the maximum C is the one selected to expand the relative terminal node of the DT to two new children nodes [31]. The above described heuristic training method for the DT was chosen as the most appropriate, since the attributes of the problem are of numerical values [33]. The following features and parameters were also considered:

(a) Dumb pruning is applied. Specifically, only dead-end nodes of 42-58% purity are pruned.
(b) The $X^2$ distribution estimator is the stop rule at a significance level of $\alpha=0.001$. [37]
(c) Terminal nodes of purity above 90% and below 10% are considered leaves and the rest dead-ends. If rules terminating with a True leaf (essentially schedules of balancing by the PC MG) do not exist in a trained DT (according to [31]), then the dead-end node of the highest purity is considered a leaf.

The complete flow of the algorithm proposed here is given in Fig. 2. The islanding of the PC MG can introduce either an excess or a deficit of a considerable amount of the total active power and is executed separately, but similarly, according to the same control framework of Fig. 2.

## 4. Validation of Decision Tree methodology for *planned community microgrids*

Two case studies were performed and are presented below. The first validates the application of the suggested methodology for planning of energy storage and the second validates the use of the methodology for energy balancing within a PC MG when a disconnection from the main grid occurs (islanding). The architecture of a PC, as this is presented in Fig. 3, is assumed [18]. The following topology is considered:

- **Block A1** consists of 160kVA domestic loads, 100kVA chiller, 200kVA various landlord loads, 200kVA CHP generation unit, 25kWp on-roof photovoltaic plant, 70kWp build-integrated photovoltaic (BIPV) installation and a 50kW battery converter.
- **Block A2** consists of 160kVA domestic loads, 100kVA chiller, 200kVA various landlord loads, 200kVA CHP generation unit, 80kWp on-roof vertical axis wind generators, 70kWp BIPV installation and 50kW battery converter.
- **Block B** consists of 320kVA domestic loads, 100kVA chiller, 300kVA various landlord loads, 75kW electric-vehicle charging station, 150kVA CHP generation unit, 40kWp on-roof photovoltaic plant, 110kWp BIPV installation and a 200kW battery converter.
- **Community Centre** consists of 1MVA mixed load, 250kVA CHP generation unit, 120kWp on-roof photovoltaic plant, 60kWp BIPV installation and a 200kW battery converter.

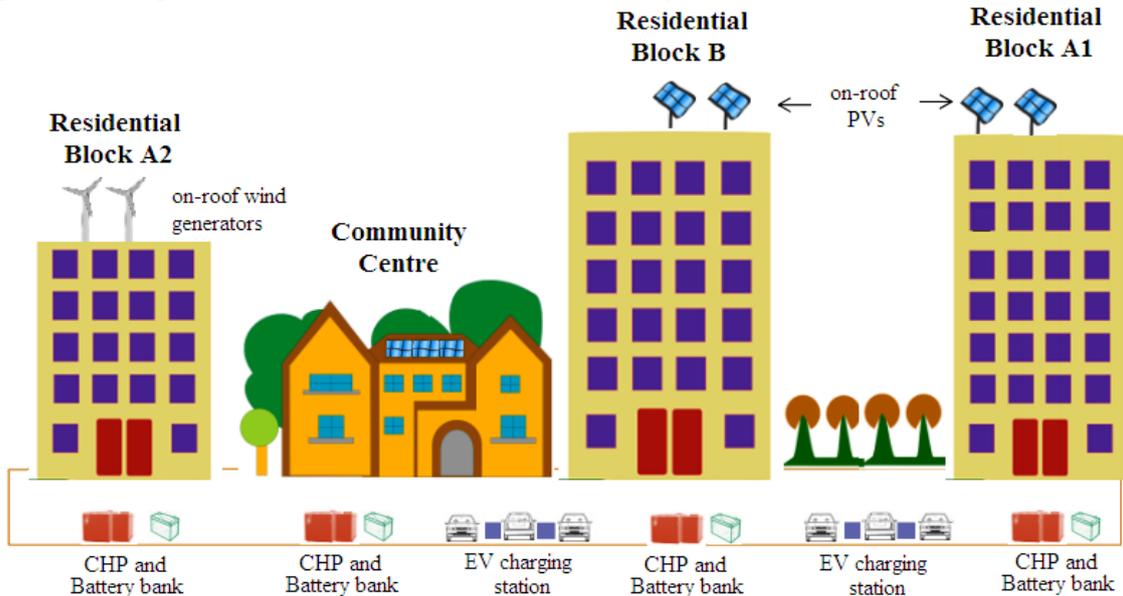

Fig. 3. Assumed planned community microgrid as of [18].

### 4.1. Storage capacity planning based on balancing deviations of a PC MG on an annual basis

Any deviation beyond ±3.3% of the nominal load of the PC MG, i.e. 90kW in this case, will have to be covered by the MG, with absolute priority that of serving the load. Since the deviations to be covered are beyond ±90kW and can reach values as low as -160kW and as high as 330kW, bins of 50kW will be used to group the dispatches. Hence, the deficits of power will be covered up to 100, 150, 200, 250, 300 and 350kW, while the excesses down to -100, -150 and -200kW.

The possibility of lower load than expected or increased power from the DG units within the PC MG, requires for either the de-loading of the available DG or absorption of the excess generation. De-loading the DG would increase the proportional energy cost of the PC MG, since the considered DG are less costly than the infeed [38]. Secondly, the availability of sufficient Renewable Obligation Contracts (ROCs) would be at risk [39]. Instead, excess power could be absorbed by storage devices. This leads to the question whether the storage system should be kept regularly at 100% State of Charge (SOC) or at a lower level so as to absorb the aforementioned excess and shift it to other times of a next day or cover for the next symmetrical event. This case study aims to address this question, examining the levels of 90% and 70% SOC. These levels will be here forth addressed as "**preferred SOC**". The 90% SOC offers a near maximum up-reserve, meaning that it can cover a high energy request when an islanding event occurs. The 70% SOC offers almost equal up and down reserves, since the lowest SOC is 20-30% for this battery technology. Hence, a storage device at 70% SOC can almost equally cover a power injection as well as a power absorption event.

Besides the nominal battery converter power, the capacity of the storage systems installed is left open to be examined through a sensitivity analysis. Therefore, four different capacity sizes at the two preferred SOC will be examined as follows, hence the capacity of the storage system is the design variable of this problem:

- **Essential backup**: only the most essential emergency loads (adding up to 300kW) can be covered for an event of a one-hour disconnection from the grid, i.e. this battery bank capacity is 300kWh. If this capacity is combined with the CHPs, it is sufficient to cover all domestic loads.
- **One-hour maximum loading of the converters**: the nominal power of the installed battery converters adds up to 500kW. Since the converters could be overloaded up to 120% for one continuous hour of operation, this battery bank capacity is 600kWh.
- **Business as usual for the whole PC MG**: A reasonable simultaneity factor of 70% among the loads is empirically assumed, thus the PC MG may have a total of approximately 2MW loads operating. In order to support this load in combination with the CHP units (800 kW) for one hour, the battery bank will need a capacity of 1.2 MWh. In this case, the converters should be upgraded, so as to avoid their overloading which was noticed in the 2030 and 2050 scenarios as analysed in [18].
- **Essential backup for the duration of a black-out**: The 300kW of emergency loads can be covered for approx. 8 hours from a storage capacity of 1.5MWh, plus some availability from CHP units.

The cost per kWh in the whole lifetime of a battery is considerably higher than that of most generating technologies. Thus, since there is no direct way to commercially exploit the availability of the storage system, the PC MG is expected to sign STOR (Short-Term Operating Reserves) contracts [40] for hourly injections of energy according to the available capacity from the batteries. When the load is high, deviation events are proportionally higher than during hours of lower loading. This assumption can also be extended to the photovoltaic generation, but cannot be considered valid for the deviations of wind generation. Combining the above events would constitute the worst case scenario.

For each battery storage scenario examined, the costs incurred in order to cover for the various deviation events is given in Table I. The results represent the highest profitability (least cost) schedule that the DT methodology could output.

*Table I. Annual costs of covering deviations of load/power from forecasted values in the PC MG for all battery storage options. ("not suitable": available capacity cannot respond to all possible scenarios)*

| Installed Capacity (kWh) | Deficit Events Cost Preferred SOC | | Excess Events Cost Preferred SOC | | Total annual cost Preferred SOC | |
|---|---|---|---|---|---|---|
| | *90%* | *70%* | *90%* | *70%* | *90%* | *70%* |
| *300* | £ 13,680.17 | not suitable | £ 2,471.70 | £ 734.34 | £ 16,151.87 | not suitable |
| *600* | £ 14,680.52 | £ 14,628.66 | £ 1,348.91 | £ 0.00 | £ 16,029.43 | **£ 14,628.66** |
| *1200* | £ 17,118.76 | £ 16,580.33 | £ 374.33 | £ 0.00 | £ 17,493.09 | £ 16,580.33 |
| *1500* | £ 19,056.84 | £ 18,608.01 | £ 0.00 | £ 0.00 | £ 19,056.84 | £ 18,608.01 |

For the case study examined, the most cost-effective option for covering the load/power deviations in the given framework, is to install a storage system of 600 kWh at a preferred SOC of 70%. A lower SOC provides more available charging-discharging cycles and, thus, a longer battery lifetime. Hence, if the battery lasts for more years, the PC MG owner/operator is paid for more yearly STOR contracts and the cost per kWh is reduced. The best storage capacity for a preferred SOC of 90% would also be 600 kWh. The utilization of the 600 kWh storage system was found to be higher than that of the 300kWh. In contrast, the utilization of the 1.2 MWh system is limited by the converter/STOR specifications assumed, hence capacity beyond 600 kWh is not really exploited. During the case study it has been observed that:

(i) RES-based DG units incurred considerable energy savings,
(ii) storage systems increase the resilience of an installation at a considerably low cost and
(iii) the application of MG control approaches can improve the electrical profile of a PC installation, by improving characteristics such as the voltage profile or power quality.

Hence, it can be concluded that the PC MG may represent a viable and profitable power system entity characterized by consistent operation, while being capable to offer power system support as an active prosumer. However, planning should be performed on a case-by-case basis and according to the installed DG capacity, the requested load and the available national or regional incentives and obligations. These parameters are particularly important for the sizing of a storage system which supports PC MG resilience.

### 4.2. Balancing load against available storage and DG of a PC MG in the event of islanding

#### 4.2.1. Methodology implementation and market framework of the islanding scenario

Balancing schedules are prepared on an hour-ahead horizon for dealing with stochastic DG and load deviations, during interconnected operation. A special case of such schedules is islanding of the PC MG due to disconnection from the power system. Based on the worst case scenario, where the PC MG consumes more power than it produces, balancing may require a compromise between using storage and load shedding. In this paper, it is proposed that this compromise is carried out through special pricing of the loads; e.g. through bilateral contracts between the PC MG operator and the loads, for emergency events. That way, the pricing itself can promote a more sensible use of loads during emergency situations. Hence, for the test system considered, it is assumed that domestic loads pay the lowest energy price, the landlord and chiller loads are priced as medium-cost customers, while the community centre pays the highest energy prices for its loads. Thus, it is more profitable for the PC MG operator/owner to serve critical residential loads, than other loads, e.g. the community centre.

The energy balancing methodology was tested against a case of islanding and the initial hourly schedule of the PC MG at the time of the disconnection is given in Table II. A summer profile has been assumed for the DGs of the system, the duration of the disconnection was set to be one hour and the installed storage capacity was assumed to be 300kWh. Nevertheless, the methodology is suitable for any length of interruption and the capacity of installed storage is not affecting the scheduling. The schedules for the five highest levels of profitability are presented in Table III. It can be seen that as the requested level of profitability increases, the methodology allows for more domestic load, while it curtails the community centre load as expected.

*Table II. Initial summer hourly schedule of the PC MG with a 300 kWh battery system.*

| DG@Bus | $P_n$ (kW) | $P_{t+1}$ (kW) | $P_{r,t+1}$ (kW) | $P_{av,t+1}$ (kW) | | |
|---|---|---|---|---|---|---|
| BIPV@2 | 70 | 42 | 0 | | | 0 |
| PV@4 | 25 | 22.5 | 0 | | | 0 |
| CHP@5 | 200 | 180 | 20 | | | 0 |
| BB@5 | 50 | 0 | 0 | 29.4 | -4.2 | SOC=90% |
| BIPV@7 | 70 | 42 | 0 | | | 0 |
| WG@9 | 80 | 56 | 0 | | | 0 |
| CHP@10 | 200 | 180 | 20 | | | 0 |
| BB@10 | 50 | 0 | 0 | 29.4 | -4.2 | SOC=90% |
| BIPV@12 | 60 | 36 | 0 | | | 0 |
| PV@13 | 120 | 108 | 0 | | | 0 |
| CHP@15 | 250 | 225 | 25 | | | 0 |
| BB@15 | 200 | 0 | 0 | 117.6 | -12.6 | SOC=90% |
| BIPV@17 | 110 | 66 | 0 | | | 0 |
| PV@19 | 40 | 36 | 0 | | | 0 |
| CHP@20 | 150 | 135 | 15 | | | 0 |
| BB@20 | 200 | 0 | 0 | 117.6 | -12.6 | SOC=90% |

*Table III. Energy balancing for islanded operation of the PC MG.*

| DT Schedule | Rule |
|---|---|
| **Top-35% profit** | if (L/l_B@21≥100) and (Dom_A2@9≥27) and (Dom_B@19≥110) and (Dom_B@18≥78) then TRUE |
| **Top-25% profit** | if and (Dom_A2@9≥35) and (Dom_B@19≥110) and (Dom_B@18≥78) then TRUE (**purity≈96%**) |
| **Top-20% profit** | if (Dom_A2@9≥35) and (Dom_A1@4≥100) and (Dom_B@19≥110) and (Dom_B@18≥102) then TRUE (**purity≈100%**) |
| **Top-15% profit** | if (Dom_A1@4≥123) and (Dom_B@18≥102) and (CC@15<13) then TRUE (**purity≈100%**) |
| **Top-10% profit** | No rule could be extracted |

#### 4.2.2. Efficiency of the Methodology

The question of whether the methodology actually yields effective cost reduction is answered through the evaluation of the efficiency of the DT tool. This approach is adopted, since there may not be an online assessment of the exact amount of balancing, hence a statistical analysis is required. Furthermore, any attempt to increase the size of the LS, in order to extract additional solutions/rules, produces only similar schedules, but no extra ones; i.e. the validity of the solutions has to be proven against the Monte Carlo methodology that generates the LS. To address the above observation, the misclassification rate ($mr$) of the DT tool will be assessed. Given any random dispatch, a misclassification occurs if the dispatch is characterized as True (False) by the LS generating algorithm and False (True) by the deepest rule of the DT corresponding to it. For example, if a dispatch of the DG units, storage and interruptible loads is described by the Top-20% profit rule, but the actual cost is above the corresponding cost threshold (i.e. less profitable), then this is a misclassification; similarly, if a dispatch yields cost less than the cost threshold of the Top-10% (i.e. more profitable), but it cannot correspond to the rule of this profit level, then this a misclassification, too, etc. The $mr$ is defined as from (2), where |TS| is the size of the test set which is usually 1/3 of the LS of the DT:

$$mr = \frac{\text{dispatches of the TS unsuccessfully classified by the DT}}{|TS|} \quad (2)$$

It should be noted that the $mr$ calculates misclassifications also from DT paths that do not represent a rule, i.e. as of Fig. 1 there are two paths describing a rule and one which is not; moreover, rule to node No. "2" is irrelevant to the methodology described (focused to rules classified as "True"). That said, mrr is defined as the mr of misclassifications concerning "True" rules, i.e. misclassifications are only dispatches that are described by "True" rules of the DT, but are "False" according to the LS generating algorithm, and dispatches that are "True" according to the LS generating algorithm, but cannot be described by a "True" rule of the DT. The average $mr$ and $mrr$ for various loading profiles of the VPP islanding scenario are given per profitability level in Table IV.

*Table IV. Average mr and mrr of energy balancing for various islanded operation scenarios of the PC MG.*

| Profit Level | Top-35% | Top-30% | Top-25% | Top-20% | Top-15% | Top-10% |
|---|---|---|---|---|---|---|
| *mr* (%) | 8.42 | 7.93 | 6.98 | 5.96 | 5.44 | 3.62 |
| *mrr* (%) | 2.81 | 1.78 | 1.40 | 1.23 | 1.09 | 0.91 |

The rate at which the proposed methodology fails to yield expected cost reduction is minimal and is decreased as the requested profitability increases; the latter as of the fact that $mr$ follows information entropy [30].

Any standard forecast errors are accounted for through the $mr$ analysis, but considerable deviations should be rectified as discussed in *Section 3.2.2* by employing lower profitability rules/schedules, which involve less contribution by stochastic sources and less domestic load consumption. For example, the profitability level of Top-15% of the islanding scenario of Section 4.2.1 (Table III) corresponds to a cost threshold of +0.0845£/kWh, while the Top-20% to +0.0793£/kWh. One of the misclassified (as "True") dispatches at the Top-15% level yields +0.0814£/kWh – due to deviation of the total PV generation from its forecasted value within the confidence interval considered, which is still more profitable than having applied a Top-20% dispatching.

Lastly, as of the methodology proposed, the difference between the cost threshold of the near global otpimum profit and the cost threshold of the schedule realized in each case, represents the cost of the forecast error.

## 5. Implementation of the DT energy balancing methodology

Due to the centralized nature of the methodology, the computational burden applied to the central processor is substantial. On an hour-ahead horizon, this would limit the number of actors that can be considered, since above a certain number of actors the computational time would increase beyond the one hour time-step. Hence, parallelization of the LS generation process would be required. In this paper, a topology of parallel processors is considered, for realizing the parallelization of the proposed methodology.

In a potential field trial of the full proposed PC MG, twenty processors would have been installed; i.e. one for every DG unit and four additional ones for the loads. The proposed PC MG was a realistic representation, but still a hypothetical system, so a proof-of-concept demonstration was set up with four processors (see Fig. 4). The hardware selected was Raspberry Pi boards, as their processing and interface characteristics were considered suitable for the particular trial. The processors in this topology generate the LS proportionally, e.g. for |LS|=1000, each of twenty Raspberry Pi processors would generate 50 elements of the LS. Then, the parts of the LS are collected by the central processor and evaluated according to the dispatch cost criterion for various profitability levels. The LS is then used to train the corresponding DTs which output the loading/shedding dispatches. The algorithm for the LS generation was developed in Java, the evaluation on the central processor in MATLAB, while the DT algorithm was developed in MS-DOS C++ and executed also on the central processor. Testing the scenario of the islanded operation of the PC MG (as in Section 4.2), the parallel topology yielded identical rules to the ones of Table III, except for minor purity (1-3%) and kW (4-5kW) differences. Each Raspberry Pi required 9 sec to generate 50 elements of the total LS. Executing the whole methodology on an Intel® Core™2 Duo@2 GHz processor (LS generation and evaluation in MATLAB, DT algorithm in MS-DOS C++) was timed at 31 sec. Hence, the computational time reduction is a significant improvement (3 times less), in addition to the redundancy and extensibility provided by the distributed nature of the system. Finally, there is no loss in performance since the parallel topology is only generating the LS of the DT in a distributed manner, using the same Java code as the centralised topology. Thus, the method is shown to be platform-independent. However, supposing that the central topology may have no limitations to the number of processing cores, if each distributed processor has less processing power than each of the cores of the central topology, then there is a proportional reduction in the execution time improvement. Nevertheless, such a comparison is only theoretical, since a central, multi-core topology suffers from cost considerations and low usability beyond the proposed methodology.

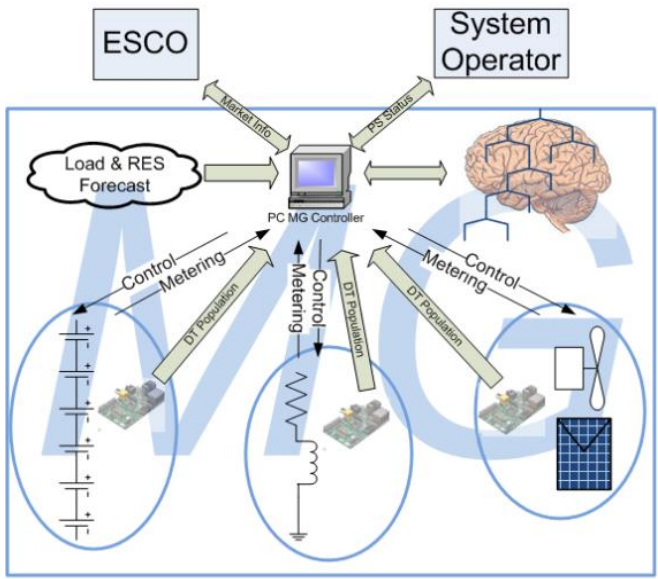 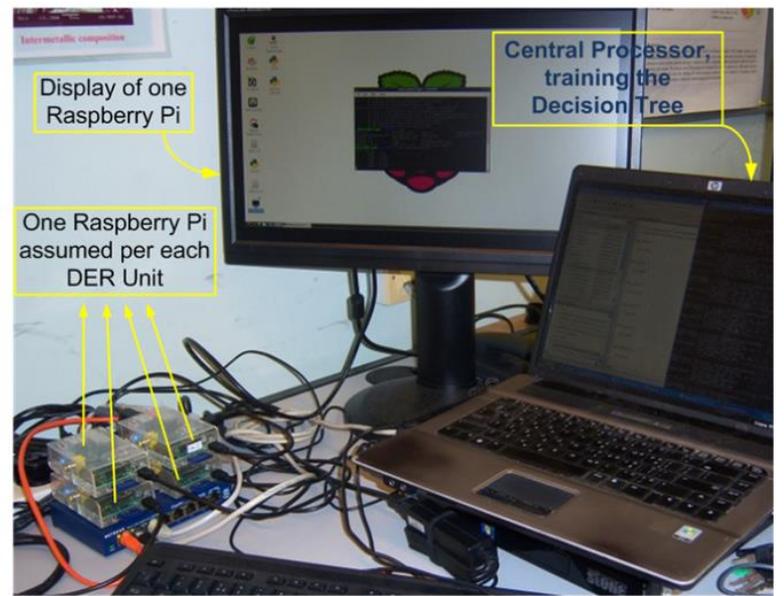

Fig. 4. Conceptual (left) and experimental (right) set-up of the methodology topology. For the experimental set-up, four Raspberry Pi processors generating one fourth of the LS each and a central processor collecting the whole LS and training the corresponding DT.

**Conclusions**

In this paper, the application of MGs on the particular residential model of PCs has been presented. The benefits of such an approach have been presented. The most important benefit is the maximization of the use of local renewable energy, but also the potential to minimize the operational cost of the PC energy supply through the optimal utilization of local energy resources. Increased network resilience is an additional benefit, especially where energy storage is present. To accomplish the above a novel DT methodology was presented, with a dual purpose:
(1) Planning for and sizing the energy storage requirements of the PC MG.
(2) Actively balancing the energy of the PC MG within an intra-day horizon.

Comparison of the effectiveness of the DT as a machine learning tool against other methodologies has been presented in [30] and is beyond the scope of this paper. The main advantage of the proposed approach against other energy balancing methodologies is that it is flexible, by means of accommodating load and RES forecasting uncertainty and by not producing a fixed dispatch, but a set of rules which if adhered to will lead to increased profit. The effectiveness of the DT methodology for this particular application has been evaluated with a case study. Cost-effective sizing of the energy storage devices has also been determined for the considered case. It was confirmed that the methodology is suitable for determining energy storage requirements, but may also be suitable for other planning studies.

The functionality of balancing the studied PC MG has also been evaluated. The DT methodology was found to effectively balance the energy in a PC MG, even in the challenging situation of an islanded PC MG system. Finally, the practical application of the methodology in distributed controllers has been tested, since distributed control may provide additional flexibility and resilience. A method for parallelization of the DT algorithm was developed and validated successfully on trial hardware, proving its feasibility.

## Acknowledgements
The work presented in this paper was supported by the Arup Global Research Challenge 2013, through the project titled "Balancing urban microgrids in future planned communities".